\documentclass[epjST]{svjour}
\usepackage[numbers,compress]{natbib}
\usepackage{graphics}
\usepackage[utf8]{inputenc}
\usepackage[T1]{fontenc}
\usepackage{amsmath}
\usepackage{amssymb}
\usepackage{hyperref}
\usepackage{changes}
\usepackage{hyperref}
\usepackage{fancyvrb}
\DeclareMathOperator*{\argmin}{arg\,min}
\DeclareMathOperator*{\argmax}{arg\,max}
	
\usepackage{lineno}
\begin{document}
\title{Trends in recurrence analysis of dynamical systems}
\author{Norbert Marwan\inst{1,2}\fnmsep\thanks{\email{marwan@pik-potsdam.de}} \and K.~Hauke Kraemer\inst{1}}
\institute{Potsdam Institute for Climate Impact Research (PIK), Member of the  Leibniz Association,
Telegrafenberg A31, 14473 Potsdam, Germany \and
University of Potsdam, Institute of Geoscience, Karl-Liebknecht-Straße 32, 14476 Potsdam, Germany}
\abstract{
The last decade has witnessed a number of important and exciting developments 
that had been achieved for improving recurrence plot based data analysis and to widen its 
application potential. We will give a brief overview about important and 
innovative developments, such as computational improvements, alternative 
recurrence definitions (event-like, multiscale, heterogeneous,
and spatio-temporal recurrences)
and ideas for parameter selection, theoretical considerations of 
recurrence quantification measures, new recurrence 
quantifiers (e.g., for transition detection and
causality detection), and correction schemes. New perspectives have 
recently been opened by combining recurrence plots with machine learning.
We finally show open questions and perspectives for futures directions of
methodical research.
} %end of abstract
\maketitle
\section{Introduction}
\label{sec_introduction}

\begin{figure}[htbp]
\begin{center}
   \includegraphics[width=.75\textwidth]{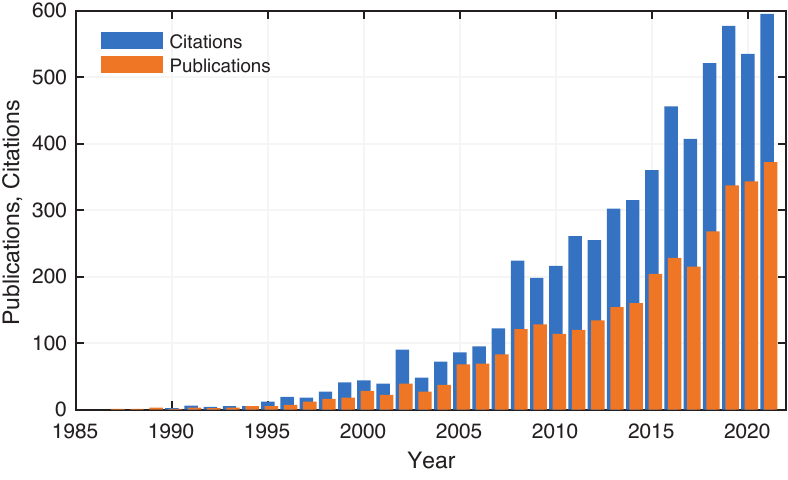}
   \caption{Number of publications and studies using recurrence plot based methods (based on 
   the database available at \cite{rpwebsite_bibliography}, May~2022)
   and citations referring to seminal studies
   as retrieved from a Web of Science search (May~2022, details in Appendix~\ref{apdx_citations}).}\label{fig_citations}
\end{center}
\end{figure}

\begin{figure}[htbp]
\begin{center}
   \includegraphics[width=.55\textwidth]{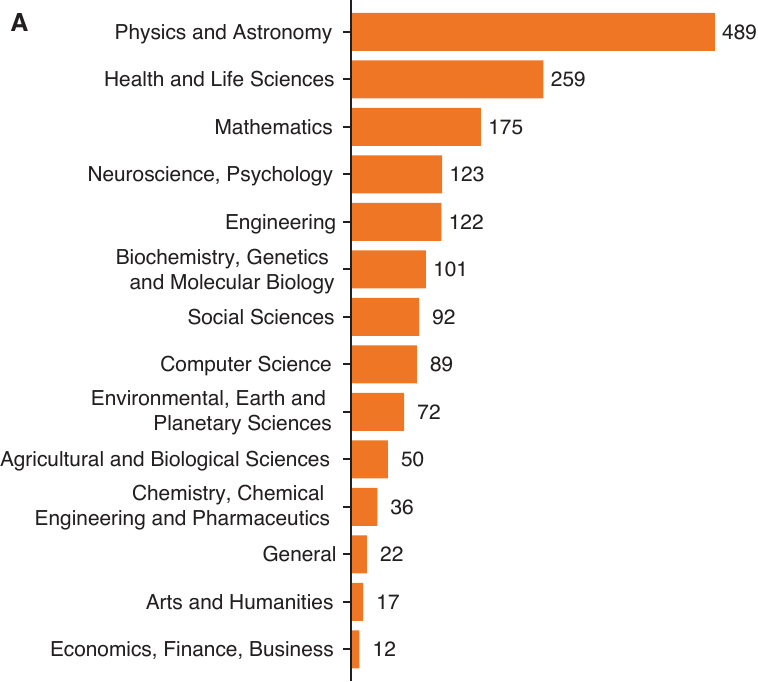}\hspace{12pt}
   \raisebox{42pt}{\includegraphics[width=.34\textwidth]{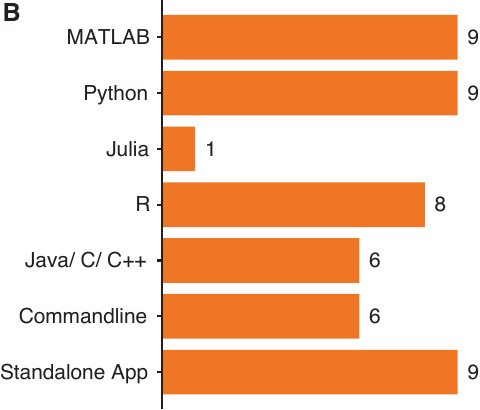}}
   \caption{(A) Subjects covered by publications using RP based methods (based
   on Scopus subject classification database in \cite{rpwebsite_bibliography}, May~2022,
   see also the notes in Appendix~\ref{apdx_subjects});
   (B) software for RP based analysis is available as standalone applications
   and as packages for the most frequently used high-level programming languages
   (based on information at \cite{rpwebsite_software}, May~2022).}\label{fig_software}
\end{center}
\end{figure}

Recurrence in dynamical systems is a fundamental feature, indicating different
types of dynamics, such as periodic, chaotic, or random variations, or predictable
and unpredictable variability. The study of recurrences in dynamical systems
by recurrence plot (RP) based methods\footnote{A RP is a matrix 
$R_{i,j}=\Theta\left(\varepsilon - \|\vec{x}_i - \vec{x}_j\|\right)$, 
representing all the times $j$ when a state at time $i$ is recurring. 
Further information on RPs and RQA can be found, e.g., in this special issue in \cite{panis2023}
or in the review \cite{marwan2007}.}, 
such as recurrence quantification analysis
(RQA) and recurrence networks (RNs) is receiving a growing interest in many
different scientific disciplines \cite{
abe2020, % 92
angus2012c, % 19
austin2019, % 70
bisi2016, % 125
bosl2018, % 227
chen2020b, % 50
curtin2018, % 245
donner2018, % 77
drews2020, % 753
eroglu2016, % 33
frasch2020, % 50
fukino2016, % 84
gandon2020b, % 56
hachijo2020, % 40
konvalinka2011, % 132
kovacs2019, % 63
lang2015, % 267
malik2020, % 369
michael2015, % 67
paxton2017, % 63
pitsik2020, % 83
shima2021, % 94
shinchi2021, % 38
twose2020, % 94
ushio2018, % 124
varni2012b, % 194
westerhold2020, % 1218
zubek2021% 65
}, %\todo{was koennte man hier kuerzen? sind sicherlich zu viele}
well represented by the increasing number of publications (Fig.~\ref{fig_citations})
and the diverse scientific disciplines these studies cover (Fig.~\ref{fig_software}A).
The increase in the number of studies citing the seminal works introducing RPs, RQA, and RNs 
\cite{eckmann87,zou2019,donner2010b,marwan2007,webber94,zbilut92} is
even stronger (Fig.~\ref{fig_citations}), which can be interpreted as a growing general popularity
of these methods not limited to the (still small although growing) community of researchers.
Nowadays, studies which are actually not using RP based methods refer to them, 
e.g., as alternative useful approaches or some kind of standard methods. 
Obviously, RP based methods are meanwhile well-accepted in
data science.

A growing number of available software is supporting this positive development (Fig.~\ref{fig_software}B).
Progress in theoretical understanding of RP analysis, GPU based computing, 
and software development in general have allowed very efficient and fast packages for 
Python and Julia (cp.~Subsect.~\ref{sect_computation}). Such packages are beneficial for working
with the challenges of big data and integrating them to machine learning approaches.
A list of software is available at \cite{rpwebsite_software}.

Big and ever growing data sets, multi-scale and spatial data, very long or very short data, data with gaps, irregular
sampling and uncertainties are challenges in many scientific disciplines. Novel
ideas and concepts are required to answer the research questions of today.
The ongoing technical developments of RP based approaches in both theoretical and practical 
domains provide tailored tools for the specific challenges. Here we have
selected a multitude of directions, ranging from computational developments, over new theoretical
insight and new recurrence definitions, to novel extensions and applications of
RP based research. It allows the interested reader to catch up on hot topics and
recent developments in RP based analysis.

\section{Trends and novel directions}

\subsection{Efficient RQA computation}\label{sect_computation}

\begin{figure}[htbp]
\begin{center}
   \includegraphics[width=1\textwidth]{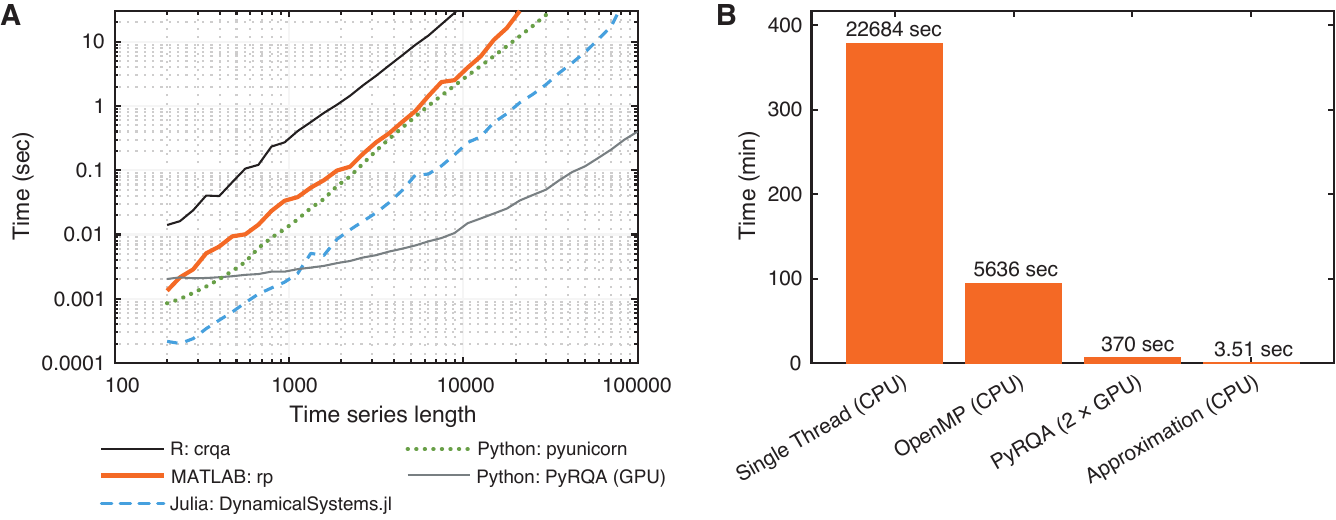}
   \caption{(A) Computation speed for recurrence plots and recurrence quantification
   measures for the R\"ossler system (details in Appendix~\ref{apdx_calctime}).
   (B) The approximative RQA allows calculation times of a few seconds
   for time series of length larger than 1 million data points,
   where standard single-thread calculations need hours (details can be found
   in \cite{spiegel2016}).}\label{fig_software_comparison}
\end{center}
\end{figure}

The recurrence matrix $R_{i,j} = \Theta(\varepsilon - \|\vec{x}_i - \vec{x}_j\|)$ 
is the basis for RP, RQA, and RN, but the calculation
of this recurrence matrix is an $N \times N$ pairwise test
(with $N$ the length of the phase space trajectory $\vec{x}_i$, $i = 1,\ldots, N$), 
thus, comes with large computational costs
in the order of $\mathcal{O}(N^2)$.
Some of the subsequent quantification (RQA measures, network measures) add a further
amount of computational complexity, usually an additional $\mathcal{O}(N^2)$. 
Therefore, long time series and big data applications require
a fast and efficient calculation of the recurrence matrix and the RP based quantifiers.
Several approaches would be possible: an efficient implementation, a parallelisation of
the computation, and approximation of the calculations.

Fast calculations can be performed, e.g., using Python, a widely used software
framework in the scientific community. The {\it pyunicorn} package \cite{donges2015pyunicorn} 
uses a very efficient implementation based on Cython and provides recurrence network measures.

Recently, the Julia language was introduced with the aim to provide a
fast and very efficient tool for scientific computations. In this line, the
Julia package {\it RecurrenceAnalysis.jl} (meanwhile integrated into {\it DynamicalSystems.jl})
was developed which also provides 
calculations for RPs and the main RQA measures \cite{datseris2018}.
The calculation of RPs and RQA measures using Julia is much faster
than comparable implementations in R, MATLAB, and Python,
in particular for longer time series $N>10,000$ (Fig.~\ref{fig_software_comparison}A).

Much shorter calculation times can be achieved by parallelising the computations.
For example, the Python package {\it PyRQA} uses a {\it divide \& recombine} approach
to distribute the computations on multi-core processors or on
an array of graphics processing units (GPUs).
The improvements can be of several magnitudes of reduced calculation time 
(Fig.~\ref{fig_software_comparison}A).

A completely different approach is using an approximation of  the
RQA measures \cite{schultz2015,spiegel2016}. 
Instead of pairwise testing the distance between all points of
phase space trajectory, the recurrences are estimated using a coarse graining of the
phase space, leading, e.g., to the recurrence rate
\begin{equation}
RR^{(m)} = \frac{1}{N^2}\sum_{i,j=1}^{N-m+1} \Theta\left(\varepsilon - \|\vec x_i^{m} - \vec x_j^{m}\|\right)
\approx  \sum_{\vec x \in X} \bigl(h_{X}(\vec x)\bigr)^2,
\end{equation}
with $m$ the current (embedding) dimension and $h_X$ the
histogram of the phase space points.
The line based RQA measures can be estimated by approximative RQA, e.g., for determinism
\begin{equation}\label{eq_approxdet}
DET^{(m)} \approx \frac{ m \cdot RR^{(m)} -  (m-1) \cdot  RR^{(m+1)} }{RR^{(1)}},
\end{equation}
(similar approximations for the other RQA measures are available, see \cite{schultz2015,spiegel2016}).
The computational complexity including the RQA measures is, thus, 
reduced to $\mathcal{O}(N\log{}N)$, resulting
in an extreme reduction of the calculation time (Fig.~\ref{fig_software_comparison}B).
However, this acceleration of RQA calculations comes with the cost of
some inaccuracies in the results.

\subsection{Alternative recurrence definitions for recurrence plots}\label{sec_definitions}

The original definition of a recurrence in phase space for creating a RP was to consider
a certain number of nearest neighbours \cite{eckmann87}. This was soon
changed to define a recurrence in terms of a thresholded distance between 
points in phase space\footnote{
A comparison of the different concepts (from the recurrence networks point of view) 
to define recurrence by the $\varepsilon$-neighbourhood 
or by the nearest neighbours approach can be found in \cite{donner2011}.}
\cite{mayerkress1989,zbilut90}. For most applications
both definitions work very well. Later, extensions were suggested to add further
criteria. Recurring points should lie on a perpendicular plane \cite{choi99} or phase
space trajectories should be parallel \cite{horai2002}, aiming to reduce the
effect of sojourn points. Order patterns are also a
very powerful extension \cite{groth2004,lu2019b}, reducing the effects of
non-stationarity, or to characterise the dynamics (cf.~Subsect.~\ref{sec_theory}).
In the last years, some additional ideas were suggested for specific research
questions.

Specific applications require tailored recurrence definitions. For the identification
of laminar regimes or to have a variance-independent distance measure, 
it can be helpful to apply the {\bf exponential function} to the actual distance
$D_{i,j} = \|\vec{x}_i-\vec{x}_j\|$
between states $\vec{x}_i$ and $\vec{x}_j$ \cite{eroglu2014,lanoiselee2021}
\begin{equation}
R_{i,j} = \Theta\left(\exp\left[
   -\frac{D_{i,j}^2}{2\lambda^2} \right]  - \varepsilon
\right).
\end{equation}
This transformation of the distances $D_{i,j}$ provides values between 0 and 1, where
1 represents the closest and 0 the longest distances. Therefore, the
thresholding is now opposite.
Such modification is, e.g., used to identify laminar regimes (cf.~Subsect.~\ref{sect_newrqa}).

\begin{figure}[htbp]
\begin{center}
   \includegraphics[width=.4\textwidth]{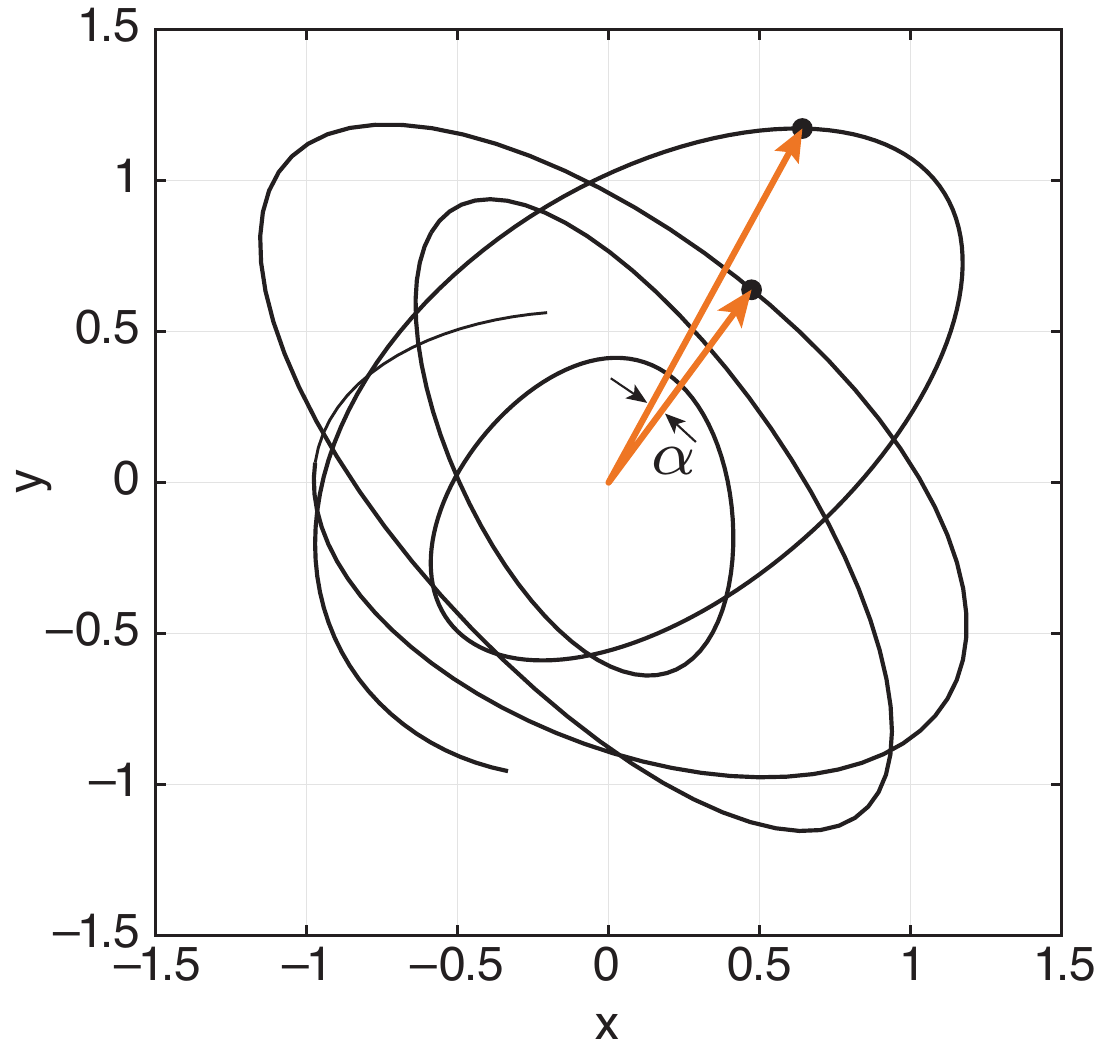}
   \caption{Instead of using the spatial distance, the angular distance, represented
   as the angle $\alpha$ between two states in phase space, can be
   used to define recurrences. Although the spatial difference between two
   points at the phase space trajectory is large, both points can be 
   considered to be recurrent because of the similar phase, indicated
   by very small $\alpha$.}\label{fig_angulardistance}
\end{center}
\end{figure}

If only 
phase differences are of interest, e.g., in material testing using ultrasonic 
signal processing, or in acoustic signal analysis, the actual amplitude should
be neglected. Here, the {\bf angular distance} is a better recurrence criterion
than the spatial distance in phase space \cite{ioana2014}
\begin{equation}
R_{i,j} = \Theta(\varepsilon - \alpha) = 
\Theta\left(\varepsilon - \arccos\frac{\vec{x}_i \cdot \vec{x}_j} {\|\vec{x}_i\|\cdot\|\vec{x}_j\|}\right),
\end{equation}
where $\alpha$ is the phase difference between both points $\vec{x}_i$ and $\vec{x}_j$.
Although the spatial difference between $\vec{x}_i$ and $\vec{x}_j$ 
can be large, they can be considered to be recurrent because of a very small 
phase difference $\alpha$ (Fig.~\ref{fig_angulardistance}). Such a recurrence criterion is particularly
useful in the analysis of ultrasonic waves for material testing or
in diagnosing atrial fibrillations \cite{brandt2016,meste2017}.

Another specific type of data, where the construction of a phase space
and measuring of distances between states at different time points might
be difficult or even impossible, are event like data. For such data,
\citet{suzuki2010} have introduced the {\bf edit distance} metric, which 
is based on transforming one sequence $S_i$ of events into another one $S_j$
(thus, the time series of events $x_i$ is segmented into short sequences of length $w$ of events
$S_i = \{x_i,x_{i+1}, \ldots, x_{i+w-1}\}$). The cost
for the minimum operations required for such a transformation is an
appropriate distance measure. The edit distance has been further extended
to better understand the parameters within this measure \cite{banerjee2021},
\begin{equation}
D_{i,j} = \min_{C}\Big\{
 \mid I \mid+\mid J \mid - 2\mid C \mid  + 
 \sum_{\left(\alpha,\beta \right)  \in C} 
\frac{1}{1+\text{e}^{-k\left(\| 
t_i(\alpha)-t_j(\beta)\|-\tau\right)}} 
\Big\},
\end{equation}
with $I$ and $J$ the set of indices of events in sequences $S_i$ and $S_j$, $C$ is the
set of {\it pairs} of event indices in $I$ and $J$, $t_i(\alpha)$ and $t_j(\beta)$ are the 
time points of the events in $S_i$ and $S_j$. The parameter $\tau$ can be used to
set the maximal delay between events when considering them as recurrent.
This edit distance can be further interpreted as a difference filter, allowing us
to construct an equidistantly sampled time series from an irregularly
sampled time series, as typical in different geoscience and astrophysics applications
\cite{eroglu2016, ozdes2022}. Consequently, a further application would even allow us
to construct RPs directly from irregularly sampled time series using such edit distance
measure \cite{hirata2015b,ozken2018}.

Such data from geosciences and astrophysics (and not only from there) has often
a certain fraction of uncertainties, e.g., from age uncertainties in
palaeoclimate archives. Instead of a series of scalar values, a time series would then 
be a series of probabilities $p(x,t)$. A recent development has combined a Bayesian approach
with RPs to derive a RP which explicitly represents also the uncertainties \cite{goswami2018}.
Instead of a binary recurrence matrix, we get a matrix with {\bf probabilities of recurrences} 
$Q_{i,j}(\varepsilon) = p\left(\|\vec{x}_i-\vec{x}_j\| < \varepsilon\right)$.
Although such representation is very helpful for data with uncertainties, the 
quantification is not as straight forward as for binary recurrence matrices.
It is still an open question how line based measures could be defined in most
reliable way (there are already some suggestions \cite{donath2019}). Nevertheless,
complex network based analysis is, of course, possible,
as it was used to identify palaeoclimate regime changes, changes in the
sea surface temperature distribution of the equatorial central Pacific, or in
financial markets \cite{goswami2018}.

An alternative for data with uncertainties are {\bf fuzzy recurrences}. Here, a 
fuzzy objective function is minimized and the fuzzy cluster membership is used to
define a recurrence \cite{pham2016} and is beneficial when working with physiological data. 
This approach can also be used for creating recurrence
networks \cite{pham2018e} and for cross recurrence analysis \cite{pham2021a}. 

Finally, for analysing spatio-temporal recurrences, we need a recurrence criterion
that considers the spatial variability in a temporal sequence of images $X(t)$.
A promising distance measure is based on the mapogram $m_{b,i,j}$ of an image $X$
which can be compared to another image $X'$ using the Bhattacharyya distance \cite{agusti2011}
\begin{equation}
D_{X,X'} = \sum_{b=1}^B\sqrt{\frac{n_b n_b'}{(\sum_b n_b)(\sum_b n_b')}} 
\sum_{i=1}^{N_i}\sum_{j=1}^{N_j} \sqrt{m_{b,i,j}m_{b,i,j}'}
\end{equation}
with $X = X(t_1)$ an image at time $t_1$, $X' = X(t_2)$ an image at time $t_2$, 
$i$ and $j$ the indices of a pixel in an image, $N_i$ and $N_j$ the size
of the image, 
$n_b$ the histogram of grey values in the image (with $B$ histogram bins), and $m_{b,i,j}$ the
mapogram (indicating the class of a pixel with respect to the histogram).
See Subsect.~\ref{sec_spatialRP} for further details on spatio-temporal recurrence analysis.

\subsection{Theoretical and parametric RQA and testing}\label{sec_theory}

The first years of RP based method development were founded by empirical findings and 
mainly lacking some theoretical background, although some connections between
dynamical properties, line lengths, and recurrence times were already 
framed 1983 by Grassberger and Procaccia \cite{grassberger83a,grassberger83b}. 
Meanwhile, several theoretical findings directly related to RPs have been achieved.

A fundamental finding was elaborated by \citet{grendar2013}, who mathematically developed
the connection between correlation sum $\mathcal{C}^{(m)}$ and the RQA measures
recurrence rate $RR^{(m)}$, determinism $DET^{(m)}$, and average diagonal line length
$L^{(m)}$. Even more important are their formulation of the asymptotic
behaviour of these measures, i.e., to which values these measures will converge when
the length of the considered data goes to infinity. 
They also show analytically that $DET$ and $L$ for Gaussian white noise do not depend
on the embedding dimension.
These considerations have been further elaborated by \citet{ramdani2016,ramdani2021},
which have further derived the analytical expressions for several RQA measures
for certain stochastic processes, fractional Gaussian noise, and correlated noise
(first analytical solutions were already given in \cite{thiel2003}).
Analytical and asymptotical expressions for RQA measures of specific random
processes are important for defining baselines for benchmarking and testing.
Moreover, the fundamental relationship between $RR^{(m)}$ and other RQA measures
can be used to define
approximative RQA measures, such as Eq.~(\ref{eq_approxdet}).

Further research has considered to derive empirical distributions for testing
serial dependencies \cite{aparicio2008,hirata2011a} and estimate $KS$ entropy
from recurrence times \cite{baptista2010}.

\begin{figure}[htbp]
\begin{center}
   \includegraphics[width=1\textwidth]{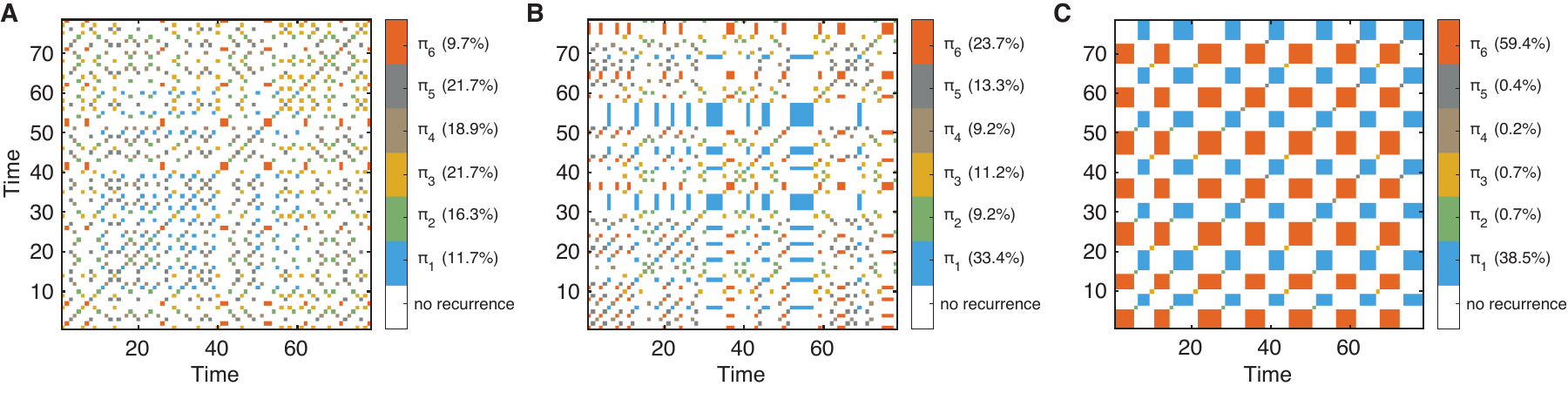}
   \caption{Order pattern RPs coloured with the corresponding order
   pattern for (A) Gaussian white noise, (B) autoregressive process of 
   1$^\text{st}$ order, and (C) $x$-component of the R\"ossler system.
   Length of order pattern $m=3$ and delay $\tau=1$, resulting in
   six different order patterns $\pi_i$, i.e., six different colours.
   The fraction of the specific recurring order pattern on all recurrences
   is provided in brackets.}\label{fig_orderpatterns}
\end{center}
\end{figure}

Another remarkable development is the use of RP based analysis to characterise
stochastic dynamics, although the original intention of RPs was to investigate
the evolution of a phase space trajectory of a deterministic dynamical system.
Based on very specific distribution of recurrence points in 
geometric pattern in the RP, the type
of the stochastic process can be determined \cite{hirata2021}.
It was shown that this approach can be used to distinguish stochastic
from deterministic dynamics and works even for short data.
Order patterns $\pi_i$ are also useful for this purpose, because some order patterns 
are very unlikely to occur for certain dynamics, called ``forbidden order patterns'' 
\cite{caballeropintado2018,lu2019b,hirata2019}. 
For example, an order pattern RP can be coloured by the specific recurring order pattern
\cite{caballeropintado2018}, providing the information about the distribution of 
occurring order patterns (Fig.~\ref{fig_orderpatterns}). 

From a more mathematical perspective, an independence test for stochastic data
was proposed, based on recurrence rate and the Cram\'er-von Mises functional 
applied to a $U$-process defined from these recurrence rates \cite{kalemkerian2020}.
The test works very well in comparison to alternative tests, like Pearson,
Spearman, or Kendall correlations, or even more advanced tests (e.g., covariance
distance).

The idea of identifying slow driving forces from time series using RPs has also been
regularly considered \cite{tanio2009,riedl2017b,hirata2020a}. \citet{riedl2017b}
combined the approach by \citet{casdagli97} with spatial RPs (cp.~Subsec.~\ref{sec_spatialRP})
to identify an external forcing on marine ecological data. A novel concept
to infer driving forces from data feeds the RP as an image-like data representation 
of the original time series into a deep learning framework \cite{hirata2020a}. The 
presented preliminary results are rather promising (see also Subsect.~\ref{sec_machinelearning} for
further combinations with machine learning approaches).

\begin{figure}[htbp]
\begin{center}
   \includegraphics[width=.75\textwidth]{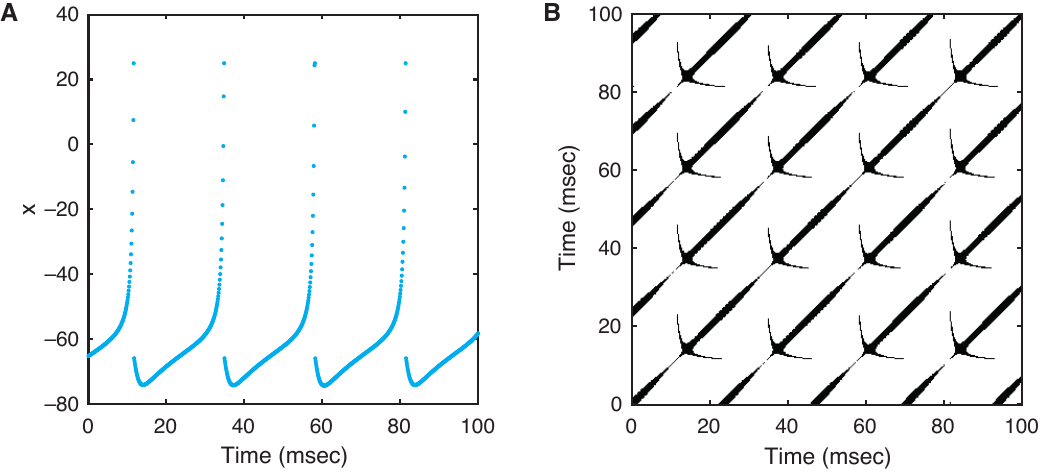}
   \caption{(A) Slow-fast dynamics derived from the Izhikevich model with
   $a=0.15$, $b=0.2$, $c=-65$, $d=8$, $I=5$, and smapling time $\Delta t=0.1$ 
   \cite{izhikevich2003}.
   (B) The very different time scales in the data of the Izhikevich model 
   cause small appendages at the 
   diagonal lines that look like sword-like structures.
   }\label{fig_slowfast}
\end{center}
\end{figure}

Few studies have investigated the small-scale structures of RPs and found links
to characteristic dynamics. We mention here two examples:
First, the shape of the block patterns in RPs is related
to specific types of intermittency \cite{klimaszewska2009}. 
Second, because of the very different time scales in slow–fast dynamics,
such dynamics causes thickening of lines or even short lines in the RP 
almost perpendicular to the main diagonal line direction
\cite{kasthuri2020} (Fig.~\ref{fig_slowfast}).

\subsection{Causal and directed relationships}

Different RP based approaches have been proposed and successfully implemented 
to detect causal and directed relationships in data. Among them are network based approaches
\cite{feldhoff2012} and joint RPs \cite{romano2004,ramos2017,peluso2020}.
The network approach uses the inter-system recurrence network, a combination of
individual RPs for both systems $X$ and $Y$ and their cross RPs
\begin{equation}
\mathbf{IR} = \left( 
\begin{array}{cc} 
   \mathbf{R}^X & \mathbf{CR}^{XY} \\
   \mathbf{CR}^{YX} & \mathbf{R}^Y
\end{array} \right)
\end{equation}
with $\mathbf{CR}^{YX} = \left(\mathbf{CR}^{XY}\right)^T$ being the cross RPs between systems $X$ and $Y$. 
Applying geometrical considerations,
the cross-transitivity coefficient (and similar cross-network measures) 
quantify how information flows between the systems, providing an indicator
on the coupling direction \cite{feldhoff2012}.

Approaches using joint RPs are closely related to mutual information \cite{marwan2007}.
The recurrence measure of dependence (RMD) is a recurrence based probability
measure similar to transfer entropy \cite{goswami2013}. Its extension is
a conditional version, the recurrence measure of conditional dependence \cite{ramos2017}
\begin{equation}
RMCD(X,Y|Z) = 
   \frac{1}{N}\sum_i\left[
     \frac{1}{N}\sum_j JR_{i,j}^{XYZ} \times 
       \log\left(\frac{\sum_j JR_{i,j}^{XYZ}\sum_j R_{i,j}^Z}{\sum_jJ R_{i,j}^{XZ} \sum_j JR_{i,j}^{YZ}}\right)
     \right]
\end{equation}
(with $\mathbf{JR}^{XYZ}$ the joint RP between systems $X, Y, Z$), 
which can be used to study indirect couplings or even causal dependencies (when considering
lagged values of one variable, e.g., $Z(t) = Y(t+\tau)$).
A similar approach is conditioning already the joint RP \cite{peluso2020}
\begin{equation}
\mathbf{CJR} = \mathbf{JR}^{XY|Z} \circ \left(1 - \mathbf{JR}^{YZ}\right).
\end{equation}
This conditional joint RP can be easily extended to more variables. $RMCD$ and $\mathbf{CJR}$ have been
shown to indicate the correct causality relationship for different kinds of challenging data \cite{ramos2017,peluso2020}.

\subsection{New RQA measures and phase space segmentation based recurrences}\label{sect_newrqa}

\begin{figure}[htbp]
\begin{center}
   \includegraphics[width=.7\textwidth]{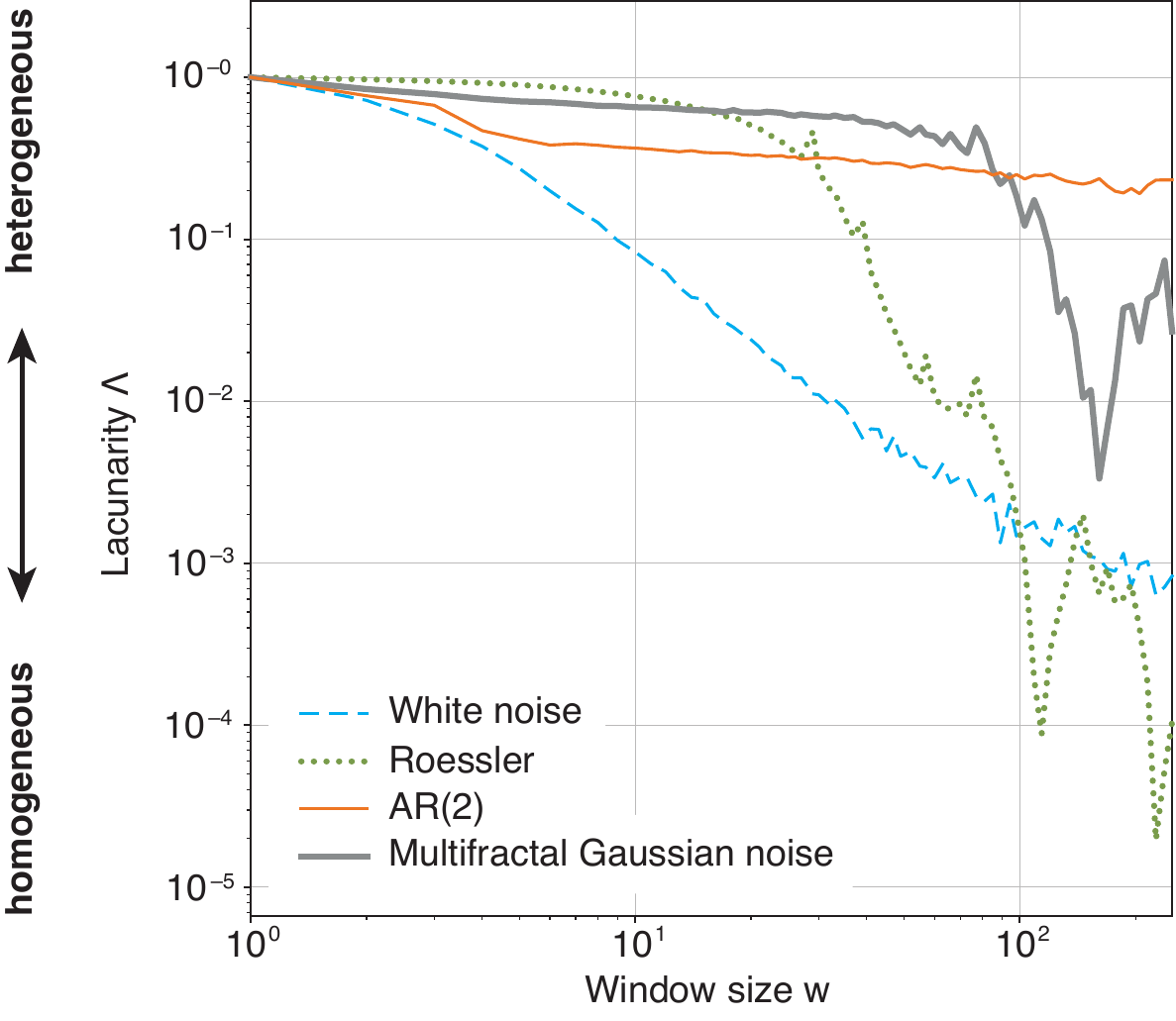}
   \caption{Lacunarity for different prototypical systems representing more homogeneous
   (white noise) and quite heterogneous RPs (AR(2) and multifractal Gaussian noise),
   as well as a RP with characteristic temporal scales (R\"ossler system).
   Technical details can be found in \cite{braun2021}.
   }\label{fig_lacunarity}
\end{center}
\end{figure}

Although the quantification of RPs has its roots in the early 1990s, there are still
some aspects that require innovative ideas for quantifying the apparently different
visual impression of RPs. Inspired by the research on fractal geometries, the
{\bf lacunarity} measure was adopted to RPs \cite{braun2021}. It characterises the homogeneity of
the RP and allows to detect characteristic time scales, such as periodicities
or extended laminar regimes (Fig.~\ref{fig_lacunarity}). 

Laminar regimes or transient trapping of states are represented in the RP by
extended blocks of recurrence points. Usually, a sliding windowing procedure is applied
to identify the changes between different dynamics.
A new measure has been suggested that can 
identify the temporal variation of transient trapping without windowing.
It is based on a {\bf block invariant measure} \cite{lanoiselee2021}
\begin{equation}
v(i) = \frac{t_\mid(i)}{t_\parallel(i) + t_\perp(i)-1},
\end{equation}
where $t_\mid(i), t_\parallel(i), t_\perp(i)$ are the geometric extensions of the 
blocks in the RP. This promising new measure was successfully applied to 
detect transient trapping events in intracellular and plasma membrane compartmentalisation.

In data analysis it can be important to identify the times of a specific
dynamical behaviour. This corresponds to a segmentation of the phase space.
\citet{beimgraben2013} have suggested several approaches to segment the phase
space into {\bf recurrence domains}, 
e.g., using the chain of transitions from one recurrence to another one, 
calling it {\bf recurrence grammars}. Such recurrence grammars are related to 
Markov chain description of the data and can be used to symbolise the RP
or to specify a new RP based entropy measure (cf.~Subsect.~\ref{sect_threshold} for an 
application of this specific entropy).

The suggested method by \citet{yang2014} goes in a similar direction, which also
segments the phase space but using a $Q$-tree segmentation. The result is a
classification of recurrences to delineate {\bf heterogeneous recurrences},
an interesting concept to reveal the fractal nature of state transitions.

\subsection{Border effects, tangential motion \& alternative RP definitions}\label{sec_corrections}

In RQA, border effects and tangential motion (sojourn points) can heavily bias diagonal line based characteristics. 
In a finite size RP these lines can be cut by the borders of the RP, thus, bias the length distribution of diagonal lines 
and, consequently, the line based RQA measures. Moreover, temporal correlations in the data, especially when
highly sampled flow data is used, noise and an insufficient embedding of the time series combined with the effect of discretization 
and an inadequate choice of parameters needed to construct the RP can cause a thickening of diagonal lines (``tangential motion'').

\begin{figure}[htbp]
\begin{center}
   \includegraphics[width=.7\textwidth]{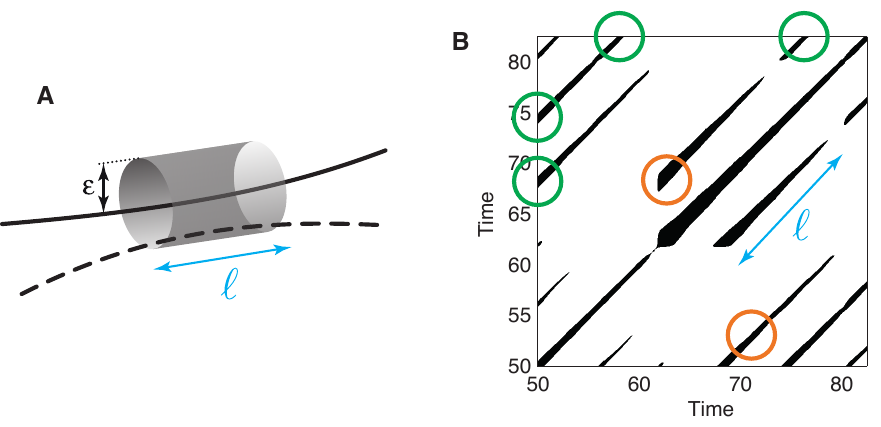}
	\caption{Parallel and close parts of a phase space trajectory (A)
	correspond to diagonal lines of length $\ell$ in a RP (B). Diagonal lines
	can be cut by the border of the RP (green circles). High sampling can cause tangential
	motion, a thickening of diagonal lines (orange circles). Modified
	after \cite{kraemer2019}.}\label{fig_border_effects}
\end{center}
\end{figure}

\begin{figure}
\begin{center}
\includegraphics[width=1\textwidth]{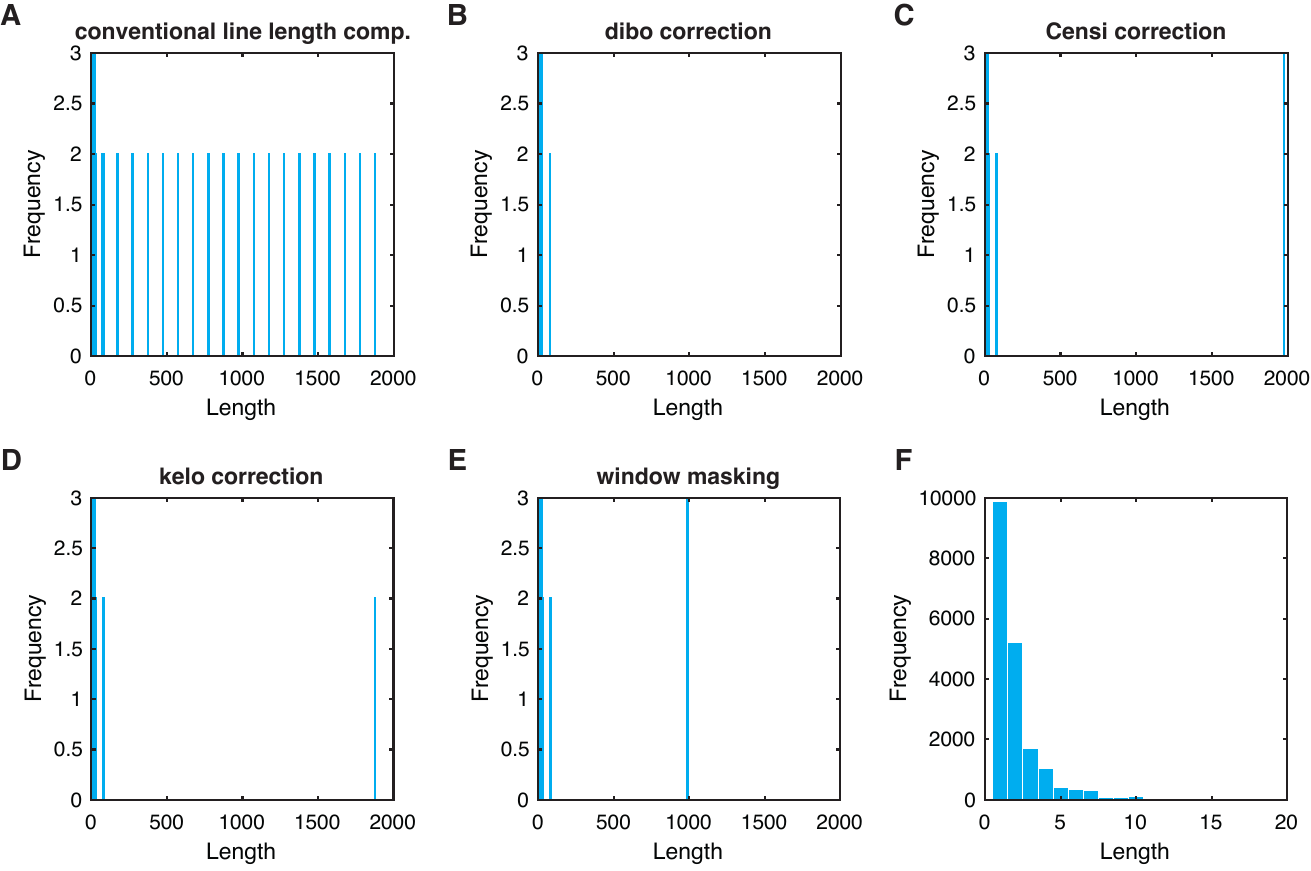}
\caption{Diagonal line length histograms of (A) the conventional line length computation and
(B) to (E) of the correction schemes proposed in \cite{kraemer2019} for a monochromatic
time-delay embedded sinusoidal with an oscillation period $T=100$ time units 
($m=2$, $\tau=T/4$).
(F) Enlargement of the histograms from panels (A) to (D), focusing on the shorter line lengths. 
A corresponding enlargement of (E) does qualitatively look the same, but with reduced
frequencies, due to the smaller effective window size. 
For a better visibility, we enlarged single bars in (B) to (E) and limited the view
to a frequency range $[0~3]$ in (A) to (E) (in (F) the full range is used).
Modified after \cite{kraemer2019}.
}\label{fig_correction_histograms}
\end{center}
\end{figure}

Both effects can have a substantial impact on certain RQA quantifiers, e.g., the diagonal line 
length entropy \textit{ENTR} (cf.~Subsec.~\ref{sec_analytical_rqa}), especially for regular dynamics. The border 
effects can be tackled in two ways. Either by a special treatment in the according histogram of any diagonal line which ``touches'' 
the RP-border \cite{kraemer2019} or by rotating the RP by $45$\textdegree (``window masking'')\cite{kraemer2019,webber2022}, in order to 
distribute the induced bias equally on all lines. For the histogram correction, we investigated in a 
previous study \cite{kraemer2019} the \textit{window masking} 
together with the ideas to either discard all border lines (\textit{dibo} correction), to only keep the longest of all border lines 
(\textit{kelo} correction), or to replace all border lines by the length of the line of identity, 
which had already been proposed by \citet{censi2004} (cp.~Fig.~\ref{fig_correction_histograms}
for a simple sinusoidal signal). 
In general, for noise free or slightly noise corrupted map data all these correction 
schemes solve the problem of the biased diagonal 
line length entropy due to lines cut by the borders of the RP.

\begin{figure}
\begin{center}
\includegraphics[width=\textwidth]{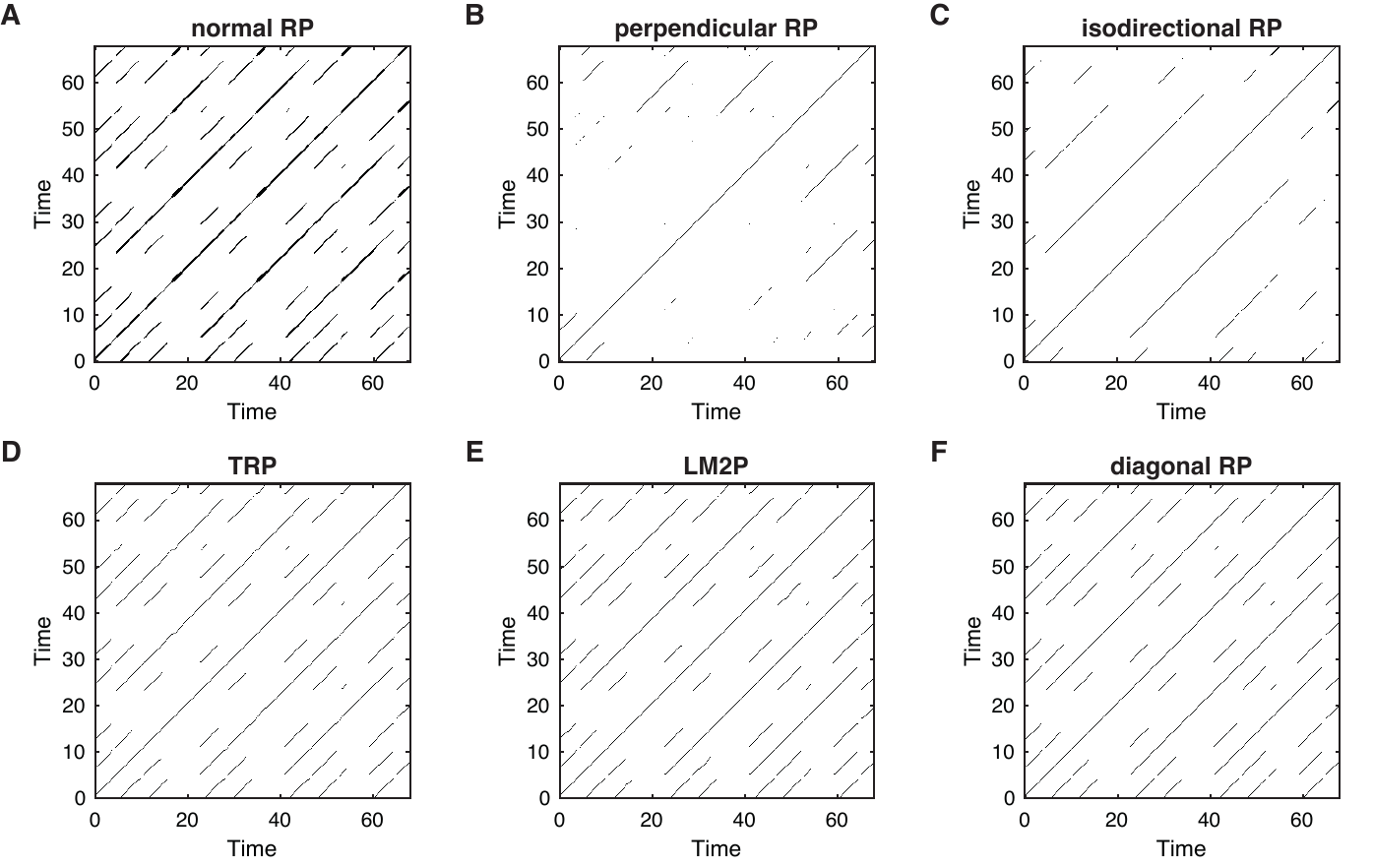}
\caption{Different approaches for avoiding the effect of tangential motion in a recurrence plot (RP), exemplary shown for the R\"ossler
system (with parameters $a=0.15$, $b=0.2$, $c=10$, sampling time $\Delta t=0.2$). 
(A) Standard RP with fixed recurrence threshold ensuring
4\% global recurrence rate as a basis to all other RPs shown in this figure. 
(B) Perpendicular RP with angle threshold $\varphi = 15$\textdegree,
(C) isodirectional RP with $T=5$ [sampling units] and $\varepsilon_2 = \varepsilon/2$, 
(D) true recurrence point RP (TRP) with $T_\text{min}=5$ [sampling units], 
which coincides with the first minimum of the mutual information,
(E) thresholded local minima approach with two parameters (LM2P) and $\tau_\text{m}=5$, and
(F) ``skeletonized'' diagonal RP.
Modified after \cite{kraemer2019}.
}\label{fig_rp_corrections}
\end{center}
\end{figure}

However, for flow data the effect of tangential motion has a much bigger influence on the entropy bias than the border effects. 
Alternative criteria of defining the RP were proposed to solve this problem 
(cp.~Subsect~\ref{sec_definitions}). The already mentioned \textit{perpendicular RP} \cite{choi99} contains only 
those points of the $d$-dimensional phase space trajectory that fall into the neighbourhood of a reference point and lie in the 
$(d-1)$-dimensional subspace (Poincar\'{e} section) that is perpendicular to the phase space trajectory at the reference point (Fig.~\ref{fig_rp_corrections}B) . 
In practice, an additional parameter is needed to account for a certain deviations of a reference point being exact on that surface 
of section. The \textit{iso-directional RP} \cite{horai2002} also promises to cope with the tangential motion, but needs two additional 
parameters (Fig.~\ref{fig_rp_corrections}C) . In this approach two points in phase space are denoted recurrent, if their mutual distance falls within the recurrence 
threshold $\varepsilon$ and the distance of their trajectories throughout $T$ consecutive time steps falls within another recurrence 
threshold $\varepsilon_2$. A further idea is the \textit{true RP} \cite{ahlstrom2006b} counts only those points to be recurrent, which first enter 
the $\varepsilon$-neighbourhood of a reference point (Fig.~\ref{fig_rp_corrections}D) . Finally, a definition of recurrences by means of local minima was suggested (\textit{LM2P} approach) \cite{schultz2011,wendi2018b}. 
In the latter approach only local minima of the distance matrix make up the RP (Fig.~\ref{fig_rp_corrections}E) . In practice a local minima detection method needs to be 
defined including an additional parameter $\tau_\text{m}$, which regulates the tolerated spacing in between consecutive minima. 

In addition to these recurrence criteria, a more geometric based approach
was proposed using a skeletonisation schema \cite{kraemer2019}. Since a ``thickened'' line consists 
of many adjacent diagonal lines, this parameter-free algorithm shrinks all ``thickened'' diagonal lines in a RP down to the longest line contained 
in such a ``thickened'' line. The result is a RP, which only consists of diagonal lines with unity width (Fig.~\ref{fig_rp_corrections}F).
Even though the true RP and the LM2P RP (Fig.~\ref{fig_rp_corrections}D, E) 
do not look to differ much from the skeletonised 
RP, in practice the computation of the skeletonised RP yields the most robust results. Together with the 
border effect corrections of the line length histograms this approach yields meaningful estimates for the diagonal line length entropy \textit{ENTR}. 
Furthermore, when computing the so corrected \textit{ENTR} for increasing minimum considered line lengths $\ell_{\text{min}}>2$ the noise level can 
be estimated and the skeletonised RP can then be used as a noise filter. The effect of these corrections on other RQA-quantifiers, including those based 
on white vertical lines (recurrence times) need to be studied as further described in Subsec.~\ref{sec_analytical_rqa}.

\subsection{Spatio-temporal recurrence analysis}\label{sec_spatialRP}

The fast development of the computational power of computers makes the
application of RPs and RQA for spatial and spatio-temporal data analysis
possible. A simple idea considered only static images and transformed the 
two-dimensional images to one-dimensional series of grey values \cite{vasconcelos2006,facchini2009}.
Unfortunately, the RQA based on this approach is influenced by the orientation 
of the spatial structures.
The more advanced approach is to compare each spatial direction of the
image, finally resulting in a RP of four or even six dimensions (for two-dimensional
or three-dimensional data, respectively) \cite{marwan2007pla}. This latter
concept is challenging because the quantification of the recurrence structures in such
dimensions is not trivial. Moreover, although all different spatial directions
are compared, objects with rotational symmetries still have an impact on the
results. Consequently, an extension was introduced by incorporating
rotations and allowing the identification of irregular circular patterns \cite{riedl2017}.
Recently, another extension was suggested to weight the grey value distances
by the distances between the pixel values \cite{chen2018}
\begin{equation}
w_{i,j} = \left( 1- \frac{\|{x}_i -{x}_j\|}{\text{range}(\vec{X})}\right)\cdot
\mathcal{D}(i,j)
\end{equation}
with $\mathcal{D}(i,j)$ a Gaussian weighted distance function of the spatial distance between
pixels $i$ and $j$. The authors have used this recurrence criterion to
construct and analyse recurrence networks of spatial data.

Spatio-temporal data such as surveillance videos or satellite data are another
interesting application field of RP based analysis. The most simple approach
would be to compare the images pixel-wise (each pixel forms the component of phase space
vector), but this would be a very
sensitive approach resulting in very low detection rates of recurrences. An
alternative would be to compare the grey value histograms. However, here
the spatial information in an image is completely lost. A powerful
approach combining both concepts was suggested by \citet{agusti2011}. 
They suggest to apply mapograms to compare images (cp.~Subsect.~\ref{sec_definitions}).
Mapograms come with a scaling factor which even allows the specific focus on 
different spatial scales that can be used in a multi-scale analysis \cite{riedl2015}.
As already mentioned, RPs based on mapograms can be used to infer the driving force 
from spatio-temporal data \cite{riedl2017}.

The identification of spatio-temporal recurrences becomes challenging when 
only a small part of an image represents a dynamical pattern. \citet{bonizzi2019}
proposed to apply a singular value decomposition (SVD) to identify the
regions of interest (i.e., the regions with some variability) and use only
the data within these regions in a regular RP. All the pixels in such a region
are considered to be the components of a phase space vector (like the simple
approach mentioned before).

\subsection{Selection of the recurrence threshold}\label{sect_threshold}

Discussions on selecting the recurrence threshold $\varepsilon$ have been
included already several times in many publications 
\cite{marwan2007,marwan2011,eroglu2014,vega2016,kraemer2018,medrano2021}.
This shows the importance of this topic, as the selection of $\varepsilon$ is a 
trade-off from having as small threshold as possible but at the same time a 
sufficient number of recurrences which strongly depends on the research question.

\begin{figure}[htbp]
\begin{center}
   \includegraphics[width=\textwidth]{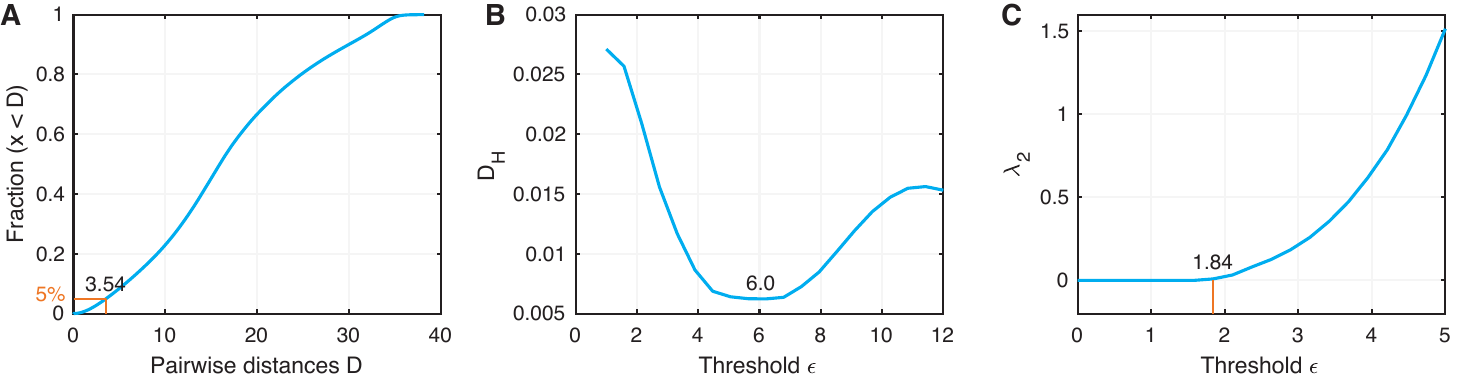}
   \caption{Finding optimal recurrence thresholds $\varepsilon$ using
   (A) quantiles, (B) topological similarity, and (C) network
   connectivity for the Roessler system (using 2,000 values of the
   $x$-component, embedded into 3-dimensional phase space using time delay
   embedding). The quantile approach (with 5\% quantile) suggests
   $\varepsilon = 3.54$, the topological similarity $\varepsilon = 6.0$,
   and the network connectivity $\varepsilon = 1.84$.
   }
   \label{fig_thresholds}
\end{center}
\end{figure}

An easy approach which helps in most cases is to use a quantile of the 
distance distribution $D_{i,j} = \|\vec{x}_i - \vec{x}_j\|$ (Fig.~\ref{fig_thresholds}A). Selecting
the threshold by using the 5\%-quantile, $\varepsilon = D_{0.05}$,
would result in a recurrence rate of 5\%. This approach provides
a robust recurrence characteristics for different embedding dimensions \cite{kraemer2018}.

Another criterion for selecting $\varepsilon$ is based on topological similarity, 
where such a value for $\varepsilon$ is selected where small changes 
$\varepsilon \pm \delta \varepsilon$ have minimal impact on the
structures in the RP. We can think about several criteria that measure
the topological similarity of RPs.
One idea for testing this is based on measuring
the Hamming distance $\Delta_H$ between the RPs thresholded with $\varepsilon, 
\varepsilon - \delta \varepsilon$, and $\varepsilon + \delta \varepsilon$,
i.e., 
\begin{equation}
\Delta_H(\varepsilon, \pm\delta \varepsilon) = 
   \frac{1}{N^2}\sum_{i,j} \bigl|R_{i,j}(\varepsilon) - R_{i,j}(\varepsilon\pm\delta \varepsilon) \bigr|.
\end{equation}
\citet{andreadis2020} have suggested to chose such an $\varepsilon$ where the
difference 
\begin{equation}
D_{H}(\varepsilon) = 
   \Bigl|
      \Delta_H(\varepsilon,+\delta \varepsilon) - \Delta_H(\varepsilon,-\delta \varepsilon)
   \Bigr|
\end{equation}
is minimal (Fig.~\ref{fig_thresholds}B). A similar idea is to consider the
RP as a RN and find network modules, again for threshold $\varepsilon$
and small deviations in the threshold
$\varepsilon - \delta \varepsilon$ and $\varepsilon + \delta \varepsilon$ \cite{vega2016}.
The first criterion is to have exactly the same number $C$ of modules, i.e.,
$C\bigl(\mathbf{R}(\varepsilon-\Delta \varepsilon)\bigr) = C \bigl(\mathbf{R}(\varepsilon) \bigr)  
=C \bigl(\mathbf{R}(\varepsilon+\Delta \varepsilon) \bigr) > 1$. The second
criterion tries to minimize the difference in the size (number of nodes) of a
given module $k$ in $\mathbf{R}(\varepsilon)$ and $\mathbf{R}(\varepsilon+\Delta \varepsilon)$,
\begin{equation}
\argmin_\varepsilon \,
\Bigl| 
  \bigl| M_k \bigl(\mathbf{R}(\varepsilon + \Delta \varepsilon)\bigr) \bigr| - 
  \bigl| M_k \bigl(\mathbf{R}(\varepsilon )\bigr) \bigr| 
\Bigr|
\end{equation}
with $M_k$ the $k^\text{th}$ module in the network and $| M_k|$ the size of the module 
(the number or nodes or phase space states in this module). This procedure
identifies such thresholds where structures in a RP do not change much for small
deviation in $\varepsilon$.

The next criterion which was suggested by several authors tries to maximize the
homogeneity of RPs. We had already seen the symbolisation based on the recurrence
grammars in Subsect.~\ref{sect_newrqa}. \citet{beimgraben2013} suggest to 
select $\varepsilon$ in a way to have the distribution of the symbols as uniform
as possible. This corresponds to a maximisation of the entropy of the
symbol distribution. A very similar approach was suggested by \citet{prado2020},
which is using local recurrence patterns of specific size (e.g., $n=2$, corresponding
to $\{R_{i,j}, R_{i,j+1}, R_{i+1,j}, R_{i+1,j+1}\}$), so called micro-states.
The criterion is to maximise the diversity of structures/patterns in the RP,
i.e., the micro-states should be equally distributed, leading to the criterion
that the entropy of the micro-states distribution should be maximal
\begin{equation}
\argmax_\varepsilon \, S\left(P(\mu)\right).
\end{equation}
As an alternative to recurrence grammars, the transition probabilities
between recurrence domains can be used \cite{beimgraben2016}. Again,
we find an optimal $\varepsilon$ where the entropy of these transition probabilities 
is maximal, ensuring equally frequent transitions between different recurrence
domains.

% prepared by threshold.m

Whereas the preceding suggestions for selecting $\varepsilon$ are mainly based on
empirical arguments and without specifying for which research question it might
work or not, \citet{medrano2021} elaborated a procedure with a deliberate theory.
The goal is to estimate dynamical invariants, like correlation dimension $C_2$ or
$K_2$ entropy. Usually, such measures should be estimated in the limit
$\varepsilon \rightarrow 0$. However, \citet{medrano2021} could show that there
will be a lower limit required, i.e., $\varepsilon \in [\beta \varepsilon_\mathrm{opt},\ \varepsilon_\mathrm{opt}]$, with $0 < \beta < 1$. Moreover, they found that
$\varepsilon$ should be selected in such a range which minimises 
the estimation errors of $C_2(\varepsilon)$ (the estimation errors when
estimating $K_2$ can also be used).

As the final approach for selecting $\varepsilon$ we mention a method
derived from complex networks. In networks, the eigenvalues of the
Laplace matrix $L_{i,j} = \delta_{i,j}\sum_j A_{i,j} - A_{i,j}$ 
(with $A_{i,j} = R_{i,j}-\delta_{i,j}$ the RN) provide information about 
the connectivity of the network \cite{eroglu2014b}. As soon as the second smallest
eigenvalue $\lambda_2$ becomes larger than 0, the corresponding $\varepsilon$
ensures that the RN will not have isolated parts, but is a connected
network (Fig.~\ref{fig_thresholds}C). This approach is related to former ideas 
of a \textit{percolation} threshold suggested 
for network based recurrence analysis \cite{donner2010b, donges2012}.

\subsection{Recurrence and machine learning}\label{sec_machinelearning}

\begin{figure}[htbp]
\begin{center}
   \includegraphics[width=\textwidth]{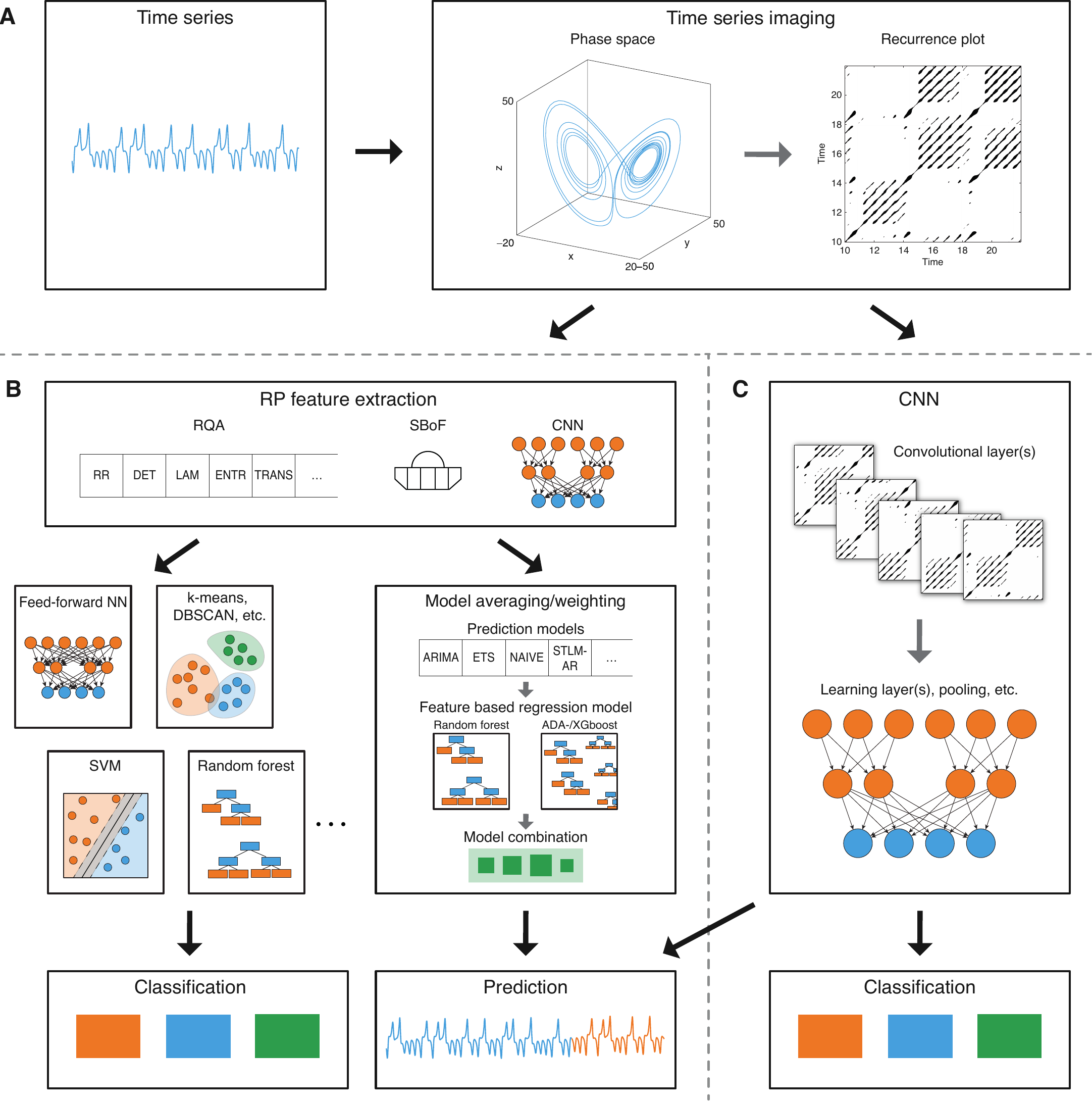}
   \caption{Simplified exemplary and schematic machine learning workflows for classification and prediction 
   using (A) the RP as an image of the time series. (B) Features of the RP can be extracted via 
   RQA yielding established features like the recurrence rate (RR),
   determinism (DET), laminarity (LAM), etc.~or via an autonomous image feature extraction algorithm, 
   e.g., spatial bag-of-features (SBoF), or pretrained convolutional neural network layers (CNN). 
   These features can be used for classification or -- in combination with 
   another regression algorithm -- for averaging/weighting of prediction models (e.g., ARIMA, ETS, NAIVE, etc.) 
   in order to obtain an optimally weighted prediction model. (C) The RP can also be used directly 
   as input to a CNN in order to classify or predict the underlying time series.   
   }
   \label{fig_machinelearning}
\end{center}
\end{figure}

Machine learning is currently a very fast-growing field. Not surprisingly
that recurrence analysis and machine learning approaches are combined
and tailored to specific research questions. Generally speaking, 
computing a RP of a time series is one way of transforming 
a sequence of data into an image, called ``time series imaging''. This transformation 
is even more complicated, when the time series gets embedded into a higher dimensional 
space beforehand (c.f., Subsect. \ref{sec_embedding_problem}). The image, i.e., the RP, can be the starting point 
of a consecutive machine learning workflow (Fig.~\ref{fig_machinelearning}A). This seems the natural way to go, since 
many machine learning tools, such as convolutional neural networks (CNN), 
were developed for image classification. Of course, other image encoding techniques such 
as Gramian angular fields or Markov transition fields instead of RPs are possible and have also 
been used \cite{estebsari2020}.

% prepared by threshold.m

Starting from the RP many different ways of setting up a ML workflow are possible 
and researchers combined several established ML-methods for classification and 
prediction tasks. First attempts started more than 15 years ago using
RQA measures as features in support vector machines (SVM) for regression
and classification purposes \cite{herrera2006,mitra2007} (Fig.~\ref{fig_machinelearning}B).
Features based on RQA measures are meanwhile frequently used
for classification purposes using SVMs, CNNs, $k$-nearest neighbour or random forest classifications
\cite{mohebbi2011,malekzadeh2021,dahmani2020a,vajiha2020,yang2018b,kirichenko2018,kirichenko2020} and the ML-toolbox offers a variety of
other methods for clustering and feature classification (Fig.~\ref{fig_machinelearning}B). 
Also other RP based quantifiers, such as based on JRPs for
synchronisation can serve as powerful features for ML classification methods \cite{yang2021d}. 
Instead of using the physically motivated RQA measures, which use certain RP-structures, such as 
diagonal lines, automated image feature extraction techniques, e.g., spatial bag-of-features (SBoF) or 
internal layer representation of a pre-trained CNN, are possible and have been used for 
forecasting in combination with another neural net, e.g., a long short term memory 
(LSTM) \cite{bi2021}. All suitable time series features can be used for forecast model averaging 
\cite[e.g.,][]{monteromanso2020}, and RP-based features appear to be a valuable complement to 
established features such as mean, autocorrelation, etc. \cite{li2020b}. The basic idea is to use the 
features in a regression model for estimating weights of a number of given forecast models, such that the 
weighted model forecast minimizes the prediction error (Fig.~\ref{fig_machinelearning}B). 
Technically, a given set of forecast models (e.g., ARIMA\footnote{autoregressive integrated moving average}, 
ETS\footnote{exponential smoothing state space model}, NAIVE, etc.) are fitted to 
the training period of each time series of the training data and produce forecasts for the corresponding 
test periods. At the same time features are computed/extracted for the training period of each time series of the training data. 
Finally the features and the prediction errors for each of the given forecast model are used to compute optimal 
weights of the models via a regression model (e.g., XGBoost \cite{monteromanso2020,li2020b}). Assuming that the time series from 
the training data and the actual data to be predicted are generated by the same process, these weights are
finally used to produce the final, improved prediction. 
    
Of course the RP can be used directly as an input for a CNN 
(Fig.~\ref{fig_machinelearning}C). Either the CNN is trained to 
classify different RPs \cite{hatami2018,nam2021,nam2021b,kirichenko2021}, or to predict time series values 
\cite{estebsari2020}. Such combinations of RPs and RQA measures with machine learning
were successfully applied for transition detection, monitoring, and
anomaly detection \cite{chen2020a,hirata2020a,seo2020,seon2021,cui2019}.

Reservoir computing (e.g., liquid state machines, echo state networks) is a specific approach
of recurrent neural networks to predict the future states of a dynamical
system based on time series without a model \cite{maass2002,pathak2018}.
RQA was used to evaluate the results of such model-free prediction
\cite{thorne2022}. But more interesting are, of course, combinations
of the learning algorithm with recurrence features. A promising approach
is to use the RQA for fine-tuning of parameters in the learning 
\cite{lee2022}.

So far, we have seen examples where RPs and RQA can help to improve
the ML applications. There are only a few studies that use ML approaches
to improve the recurrence analysis. One idea is to use a learning algorithm
to classify the RP with respect to the underlying dynamics \cite{garciaceja2018}.

\section{Perspectives for future research}

The trends in the methodological developments of RP based methods and their
applications show the perspectives for future research.

\subsection{The embedding problem}
\label{sec_embedding_problem}

The RP considers recurrences of the trajectory $\{\vec{x}_i\}_{i=1}^N$ (with $\vec{x}_i=\vec{x}(t_i)$) 
of the considered dynamical system's phase space. 
However, in most applications, $\vec{x}$ cannot be measured directly or completely, and only a 
subset of observables is available. In such cases, $\vec{x}$ must be 
reconstructed from the measured observables. All of the numerous published methods for reconstruction of the phase space 
(e.g. \cite{kraemer2021,casdagli1991,gibson1992,uzal2011,nichkawde2013}) 
introduce a certain number of parameters on which, 
consequently, the calculated RP and the RQA depend. 
This is a current field of research with the aim in automatising this process and making it robust with 
respect to a subsequent recurrence analysis (e.g., recently introduced 
PECUZAL embedding algorithm \cite{kraemer2021}). However, it has been shown that the optimization of 
embedding parameters does depend on the actual research question \cite{kraemer2022}, 
like computing dynamical invariants or prediction 
\cite{judd1998,holstein2009,bradley2015,Garland2015,wendi2018}.

The PECUZAL algorithm can occasionally suggest contradictory embedding parameters. For example, 
the logistic map is clearly a deterministic system and, therefore, the used
test statistic ($L$-statistic, \cite{uzal2011}, related to the false nearest 
neighbor statistic \cite{kennel1992,kennel2002,cao1997,hegger1999,krakovska2015}), 
should recommend an embedding with dimension $m>1$. However, in chaotic regime, PECUZAL suggests no 
embedding and treats the input as a stochastic signal. 
For other maps, e.g., the Ikeda or H{\'e}non map, this is not the case. This does 
not seem to be a problem of the specific test statistic or the PECUZAL algorithm.
When running the 
``stochastic indicator'' proposed by \citet{cao1997}, it also values the chaotic logistic map as a 
stochastic source and does not suggest any embedding. A similar problem arises when 
analyzing map-like data in a geoscientific context. These time series are often interpolated and despite 
their inherent non-stationarity we should be able to embed small pieces with approximately 
constant parameters. In many cases, ranging from drill core data under a certain age model to climate 
index data such as the Southern Oscillation Index (SOI) and to Earth system models of 
intermediate complexity (EMICs), PECUZAL does not suggest any embedding and also other stochastic 
indicators would treat the signals as stochastic. 

Therefore, the following research questions should be addressed in the future:
(1) How does interpolation affect the estimation of the embedding parameters?
(2) 
How does the sampling resolution affect the estimation of the embedding parameters 
(flow-like vs.~map-like data)?
(3) 
Countless real world processes can be described by a Langevin equation. Yet, to our best knowledge there 
is no study which systematically investigates the embedding of systems described 
by such a stochastic differential equation.
(4)
The impact of the embedding on phase space based causality measures such as convergent cross 
mapping \cite{Sugihara2012} (and its extensions), joint recurrences \cite{ramos2017,zou2011},
or recurrence networks \cite{feldhoff2012} have only been investigated briefly  
\cite{kraemer2022,Schiecke2015}. Since causality analysis is of great interest 
in many scientific (and commercial) fields, more thorough research on this topic is of high importance.

\subsection{Recurrence definitions}

In order to visualise recurrences of a phase space trajectory, we
have to define what {\it recurrence} or actually {\it similarity} of states
actually means in the context of the current research question. 
For this purpose, dynamical similarity is mostly measured in terms 
of some metric distance $D_{i,j}=\|\vec{x}_i-\vec{x}_j\|$ defined in the underlying system's $d$-dimensional phase space. However, specific data or research questions can require modifications
of this similarity measure (Subsect.~\ref{sec_definitions}). 
Depending on the further growing field of recurrence
analysis and ever new applications as such as in machine learning, 
novel similarity measures or metrics will
be required, such as for comparing field data and spatial patterns,
or time series with uncertainties and gaps \cite[e.g.,][]{abid2021}.

\subsection{Recurrence threshold}
\label{sec_recurrence_threshold}

Even though many studies (Subsect.~\ref{sect_threshold}) have considered the question of how to objectively 
find an optimal recurrence threshold, this is still not yet answered satisfactorily. In most 
applications where comparisons or relative results are of interest, 
fixing the recurrence rate at a certain value and adjusting the threshold accordingly \cite{kraemer2018} 
will be appropriate. For other research questions (such as characterising the
specific dynamical properties), a reasonable, very specific threshold should be selected.
Although several ideas for an objective selection were 
suggested \cite{vega2016,medrano2021,andreadis2020,prado2020,eroglu2014b,matassini2002a}, 
they are mainly based on heuristic ideas and the used criteria miss an objective physical foundation
(e.g., why should be a topological invariance desirable, why should be the
diversity of structures in RP maximised, why should be the recurrence network connected?).
An objective criterion should either minimise the estimation error for the dynamical
invariants \cite{medrano2021} or be a trade-off of maximising the number and length of diagonal line 
structures and minimising the threshold value itself. Besides the specific selection
criterion, a systematic overview generalising typical applications of RP based analysis
and best suited threshold selections would be helpful in particular for new
users of the method.
%
%\begin{itemize}
%\item
%Similar to the approach of using the maximum entropy of microstates for selecting the optimal recurrence 
%threshold, one could think of utilizing our proposed skeletonization scheme 
%\cite{kraemer2019} for that purpose. For an increasing threshold one counts the number of diagonal lines 
%contained in the corresponding skeletonized RP. If the threshold is 
%too high, distinct lines (\textit{distance ranges} in the jargon of \citet{kraemer2019} merge and the 
%skeletonization algorithm will reduce these merged distance ranges to a 
%single diagonal line. Therefore, the optimal threshold would maximize the number of diagonal lines.
%\end{itemize}

\subsection{Analytical RQA}
\label{sec_analytical_rqa}

The analytical explanation of various RQA measures has made great progress in recent years 
(Subsec.~\ref{sec_theory}).
However, the relation between the line structures in the RP and
dynamical invariants has not yet been satisfactorily answered.
For example, as shown in \cite{kraemer2019}, the analytically derived relation between 
\textit{ENTR} and $K_2$ \cite{march2005} does not 
yield meaningful results for real-world time series -- neither for the border effect corrected \cite{kraemer2019}, nor for the uncorrected \textit{ENTR}.

\begin{figure}[htbp]
\begin{center}
   \includegraphics[width=.65\textwidth]{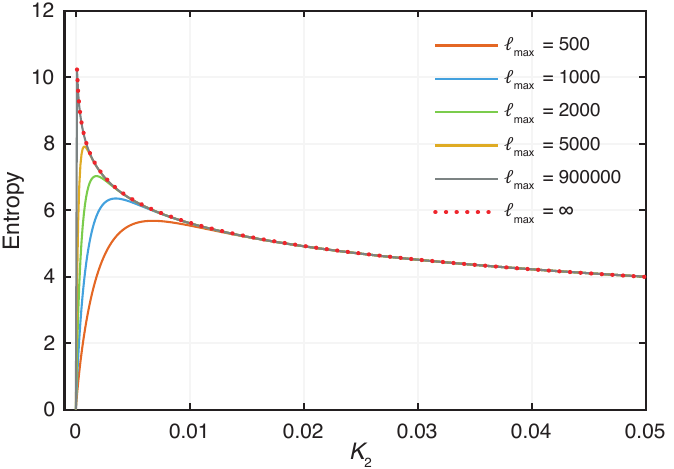}
   \caption{Theoretical values of the diagonal line length entropy, Eq.~\eqref{eq_entropy_approx}, for different upper
   limits of the maximum encountered diagonal line $\ell_\text{max}$. The case $\ell_\text{max}=\infty$ corresponds to the analytical 
   expression in Eq.~\eqref{eq_entropy_theo}.
   }
   \label{fig_theoretical_entropy}
\end{center}
\end{figure}

For an analytical expression of \textit{ENTR}, we use the limit of infinitely large RPs, 
thus, infinitely long diagonal lines $\ell_{\max}=\infty$ \cite{march2005}:
\begin{equation}
{ENTR}_{\text{theo}} = -\sum_{\ell=\ell_{\min}}^{\ell_{\max}} p(\ell) \ln p(\ell),
\label{eq_entropy_approx}
\end{equation}
with $p(\ell)$ being the theoretical probabilities of observing a line of length $\ell$.
By using the scaling property of the correlation sum with the correlation entropy $K_2$ \cite{grassberger83a}, $p(\ell)$ can be expressed in terms of $K_2$, 
$p(\ell) = \left( 1-e^{-K_2} \right) e^{-K_2(\ell-1)}$,
thus, we find a theoretical expression for the diagonal line length entropy \cite{march2005}
\begin{equation}
{ENTR}_{\text{theo}} = K_2 \left( \frac{1}{\gamma} - 1 \right) - \ln{\gamma},
\label{eq_entropy_theo}
\end{equation}
with $\gamma = (1-e^{-K_2})$. For increasing $K_2$, \textit{ENTR} will decrease (Fig.~\ref{fig_theoretical_entropy}). Eq.~(\ref{eq_entropy_theo}) holds only in the
limit of $N \rightarrow \infty$, but in real world applications, we have finite
time series lengths, i.e., $N \ll \infty$, thus, the upper limit in the
sum of Eq.~(\ref{eq_entropy_approx}) is $\ell_{\max} \ll \infty$. This results
in significant deviations of \textit{ENTR} from the theoretical value in the weak 
chaotic regime, $0 \leq K_2 \leq 0.01$ (Fig.~\ref{fig_theoretical_entropy})
and is especially important for real world applications with data set lengths $<5,000$. 
In principle, it should 
be possible to get the ``right'' approximation by considering the length of the available data. 

% Figure produced by "data/Theoretical diagonal line entropy/approximate_diagonal_entropy.m"

\begin{figure}[htbp]
\begin{center}
   \includegraphics[width=.65\textwidth]{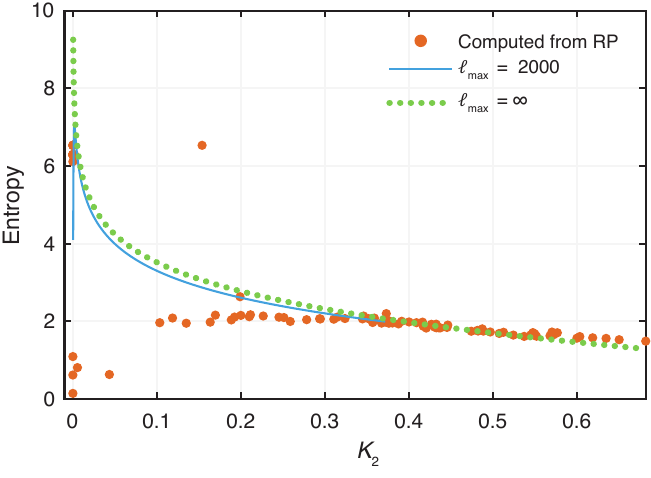}
   \caption{Relationship between \textit{ENTR} and $K_2$ for the logistic map, calculated
   from time series of length $N=2,000$ embedded in two dimensions with unity lag, a 
   fixed recurrence threshold $\varepsilon=0.05$, a minimum line length $\ell_\text{min} = 2$,
   and the \textit{kelo} correction applied. For lower 
	choices of the recurrence thresholds the graphs looked similar and only for higher thresholds the agreements with the expected values got 
	worse.
	}
   \label{fig_Logistic_entropy}
\end{center}
\end{figure}

However, when calculating \textit{ENTR} from RPs and comparing it with the approximated values
derived from Eq.~(\ref{eq_entropy_approx}), we find strong discrepancies
in particular for $K_2 < 0.3$
(Fig.~\ref{fig_Logistic_entropy}). These differences remain also for many different 
parameter settings (e.g., very small $\varepsilon$) and different systems (the correction for 
border effects, as briefly discussed in Subsec.~\ref{sec_corrections}, 
also do not improve this negative result). Of course, the main problem when 
using real world data or even model flow data is that it is not trivial to estimate 
$K_2$ properly, which could be the potential reason for such strong deviation.  
But even for a very simplistic system like the logistic map, where we can 
analytically compute $K_2$ by the positive Lyapunov exponent 
$\lambda(r) = \bigl\langle \log(|r-2rx|) \bigr\rangle$, the theoretical relationship 
as visible in Fig.~\ref{fig_theoretical_entropy} cannot be approximated. 

% Figure produced by "data/Theoretical diagonal line entropy/entropies_logistic_map.m"

Addressing the following questions would be helful in 
order to make advances in transition and bifurcation detection as well as classifying regimes:  
(1) Further elaborate the relationships between structures in RPs (diagonal
and vertical lines, recurrence times) and dynamical invariants; compare
the different estimations based, e.g., on line length distributions \cite{faure98,thiel2004a},
recurrence rate \cite{grassberger83a}, 
recurrence entropy \cite{march2005}, or
recurrence times \cite{baptista2010}.
(2) Investigate the sampling effect on these relations \cite{grassberger1988}
and clarify why some of these relations (such as the $ENTR-K_2$-relation, Eq.~(\ref{eq_entropy_theo}) and similar the $DET-K_2$-relation) do not match observational data.
(3) 
A thorough study on the impact of the correction schemes \cite{kraemer2019} (see Subsec.~\ref{sec_corrections}) on the estimation 
of dynamical invariants is needed. 
%\item
%Furthermore, the impact of these correction schemes on the $\tau$-RR and its corresponding inter spike 
%spectrum \cite{kraemer2022b} should be investigated.

\subsection{Significance tests for RQA}
\label{sec_significance_of_RQA_results}

In cases where the experimental design allows the acquisition of distributions of RQA characteristics, it 
is possible to make statements about the significance of the results. In most passive experiment setups, as it is often the case 
in medical applications, astrophysics, or geoscience, this is not possible. 
Hypotheses testing on observations of a system should then be performed
using known test models (which correspond to the null). Recent theoretical
work which derived the theoretical values for RQA measures of specific
systems (mainly stochastic systems) will help in evaluating and benchmarking
results \cite{ramdani2016,ramdani2021}. However, this approach works only
for specific null-hypotheses (e.g., testing against noise).
For more general hypothesis testing, we will rely on \textit{surrogate data}, an 
appropriate Monte Carlo sample of the underlying data for a given null hypothesis. 
Surrogates are generated by keeping characteristics of the observed system related to the null, but induce 
randomness at the same time (constrained randomisation) \citep{Schreiber2000}. 
In the context of recurrence analysis this translates into the question of how to construct surrogate phase space trajectories, 
distance or recurrence matrices, which are consistent with the 
null. It would, thus, be beneficial to construct surrogates of phase space trajectories in order to obtain 
distributions of corresponding RQA statistics, which could then be used for 
statistical testing. 

A promising method is using {twin surrogates}
\cite{thiel2006b}, which constructs 
surrogates from 
(1) identifying twins in the phase space trajectory 
(points which share the same neighbourhood) and 
(2) randomly jump to one of the possible futures of the existing twins. The drawback is, of course, that for 
proper statistical testing we would seek around $1,000$ surrogates or more 
and in the described method this number is determined by the total number of twins, which is a property of 
the data and is often too small. 

Another idea for line based RQA statistics in the running window approach
for transition detection
is based on bootstrapping line structures \cite{marwan2013}. 
To estimate the unknown variance of the diagonal line 
length distribution of a RP, surrogate line 
length distributions are bootstrapped from the cumulative line 
length distribution of all windows. Although this approach is working 
well in most cases, the resulting confidence intervals are sensitive to 
the number of bootstrapped lines. There is no objective way to determine this number,
because the number of lines can vary between the windows.

Thus, there is still an urgent need for robust methods that construct 
RP surrogates, which preserve the basic properties (correlation structure) 
of, e.g., the RP or the underlying state space trajectory. This would affect 
all existing RQA measures and would allow to make statements about the statistical relevance a measured RQA statistic has, even in passive experiments 
with single runs.

\subsection{Machine learning combined with recurrence analysis}
\label{sec_perspectives_ML}

Machine learning (ML) approaches become more and more accepted and used also
in complex systems science. RPs and RQA are already used as features
in ML applications mainly for classification purposes, but also automated feature extraction 
methods are increasingly used (Subsect.~\ref{sec_machinelearning}). 
The main question here is whether the RQA features, some of which have a relationship to 
dynamical invariants (i.e., have physical meaning), are a useful preprocessing step before 
applying a particular ML method for classification or prediction. Or whether suitable 
image feature extraction methods, such as CNNs are the way to go. Certainly, the computation of 
RQA features does not depend on too many free parameters and, thus, does not require any 
additional hyperparameter optimization or training. This is an important point, as multiple stacked 
ML methods easily become unmanageably complex models that are potentially prone to overfitting and additionally 
require a large amount of (stationary) training data. For the direct application of CNNs to the 
time series image (Fig.~\ref{fig_machinelearning}C) a sound study is also needed 
examining the difference between feeding a RP or the unthresholded distance matrix.

New directions in using ML approaches are time series based predictions using reservoir computing,
which might benefit by applying concepts from RPs and the according recurrence networks.

The future developments with respect to ML and RPs will see further 
cross-fertilisations. For example, ideas of time series imaging used for ML based
classifications such as gramian summation fields and markov transition fields
\cite{dias2020,esmael2021} could provide new definitions for recurrences.

\section{Conclusions}

Methodical research on recurrence plots (RPs) and recurrence quantification analysis (RQA)
is still a lively field. The last years have revealed a number of
important new solutions for specific research questions, but also gave
some answers to more general challenges in RP based data analysis.
Nevertheless, there are still further open ends and directions which
should be considered in the future.

\section*{Acknowledgements}

Supported by the DFG, projects
MA4759/9-1 (Recurrence plot analysis of regime changes in dynamical systems) and
MA4759/11-1 (Nonlinear empirical mode analysis of complex systems: Development of general approach and application in climate).

\section*{Data and Code}

Code used to prepare the figures is available via Zenodo
\href{https://doi.org/10.5281/zenodo.6623542}{doi.org/10.5281/zenodo.6623542}.

\appendix
\section{Appendix}

\subsection{Citations}\label{apdx_citations}

To measure the number of citations per year for the basic, most cited works, a 
search query was placed at Web of Science (2022-05-04) using the
DOIs of the paper \cite{eckmann87,zou2019,donner2010b,marwan2007,webber94,zbilut92}.
The search query for DOIs was

\begin{Verbatim}[frame=single]
10.1016/j.physrep.2006.11.001 or 10.1209/0295-5075/4/9/004
or 10.1152/jappl.1994.76.2.965 or 10.1016/0375-9601(92)90426-M
or 10.1016/j.physrep.2018.10.005 or 10.1088/1367-2630/12/3/033025
\end{Verbatim}

The search results are available via \\
\href{https://www.webofscience.com/wos/woscc/citation-report/bd5b7e1a-63b3-4345-9626-215d23f8e7e1-35998ece}{https://www.webofscience.com/wos/woscc/citation-report/bd5b7e1a-63b3-4345-9626-215d23f8e7e1-35998ece}

\subsection{Subjects}\label{apdx_subjects}

The database of publications ($N=3,618$ by May~2022) on or using RP based methods \cite{rpwebsite_bibliography}
is used to retrieve the Scopus subjects of them. This has been performed using 
a Python script to get this information via the Altmetric web service. Some of the
Scopus subjects were summarised because of significant overlap (Tab.~\ref{tab_mergedsubjects}).
A publication can cover multiple Scopus subjects, therefore, the information in Fig.~\ref{fig_software}B 
does not mean exclusive subjects per publication and the total sum of presented subjects 
does not correspond to the total number of publications.

\begin{table}[htbp]
\caption{Merged Scopus subjects}
\begin{center}
\begin{tabular}{ll}
{\it Summary subject}			&{\it Subjects included}\\
\hline
{\bf Health and Life Sciences}	&Health Sciences\\
							&Medicine\\
							&Health Professions\\
							&Nursing\\
							&Life Sciences\\
\hline
{\bf Physics and Astronomy}		&Physics and Astronomy\\
							&Physical Sciences\\
\hline
{\bf Economics, Finance, Business}&Decision Sciences\\
							&Business, Management and Accounting\\
							&Economics, Econometrics and Finance\\
\hline
{\bf Engineering}					&Engineering\\
							&Energy\\
							&Materials Science\\
\hline
{\bf Chemistry}					&Chemistry\\
							&Chemical Engineering\\
							&Pharmacology, Toxicology and Pharmaceutics\\
\hline
{\bf Biochemistry, Genetics and Molecular Biology}&Biochemistry, Genetics and Molecular Biology\\
							&Immunology and Microbiology\\
\hline
{\bf Neuroscience, Psychology}	&Neuroscience\\
							&Psychology\\
\hline
{\bf Environmental, Earth and Planetary Sciences}&Environmental Science\\
							&Earth and Planetary Sciences\\
\hline

\end{tabular}
\end{center}
\label{tab_mergedsubjects}
\end{table}

\subsection{Measuring the calculation time for recurrence analysis}\label{apdx_calctime}

We measured the calculation time for creating a RP and calculation of the standard RQA measures
depending on data length $N$
for the R\"ossler system with the standard parameters ($a=0.25$, $b=0.25$, and $c=4$) \cite{roessler1976}
and a sampling time of $\Delta t = 0.05$.
We used only the $x$-component of the R\"ossler system after removing the first 1,000 
points as transients and applied a simple time delay embedding with $m=3$ and $\tau=6$. 
The RP and RQA calculations were implemented in MATLAB (Version R2022a), R (Version 4.0.2), 
Julia (Version 1.6.4), and Python (Version 3.8.8). 
For MATLAB we used the {\it rp} code v1.1 provided by \cite{marwan2021zenodo}, 
for R the {\it crqa} package v2.0.2 \cite{coco2014}, 
for Julia the package {\it DynamicalSystems.jl} v1.4.0 (RecurrenceAnalysis v1.5.2) \cite{datseris2018},
for Python the {\it pyunicorn} v0.6.1 package \cite{donges2015pyunicorn}, 
as well as the
{\it PyRQA} v8.0.0 package \cite{rawald2017}. The {\it CRP Toolbox} for MATLAB was not used,
because the implementation is interwoven with a graphical user interface and, thus, 
the new rendering engine of MATLAB is strongly interfering and slowering
the calculations since its introduction
in 2014 \cite{crptoolbox}. 

The recurrence analysis was performed on the time series obtained from the 
R\"ossler system with growing length,
starting with $N=200$, increasing in steps to provide equidistant points along the
$x$-axis in a log-log plot. The increase of length was stoped when it exceeded 100,000 or when 
the calculation time
exceeded 30~sec (i.e., final time series had lengths
200, 237, 282, 335, 398, 473, 562, 668, 794, 944, 1,122, 1,334, 1,585, 1,884, 2,239, 2,661, 3,162, 
3,758, 4,467, 5,309, 6,310, 7,499, 8,913, 10,s593, 12,589, 14,962, 17,783, 21,135, 25,119, 29,854, 
35,481, 42,170, 50,119, 59,566, 70,795, 84,140, and, 100,000). 
For each selected length, the calculation time was measured
5 times and then averaged.

The calculations were performed on a 2.3~GHz Quad-Core Intel Core i7 with 16GB RAM, except
the calculations using the {\it PyRQA} package, which were performed on a Nvidia 
GPU Tesla V100 with OpenCL 1.2.

\bibliographystyle{unsrtnat}
\bibliography{rp,misc,hauke}

\end{document}